\documentclass[english]{article}
\usepackage[T1]{fontenc}
\usepackage[latin9]{inputenc}
\usepackage{textcomp}
\usepackage{amsmath}
\usepackage{amssymb}
\usepackage{esint}
\usepackage{babel}
\usepackage{mathrsfs}
\begin{document}
\title{The geometric tensor for classical states}
\author{A. D. Berm\'udez Manjarres\thanks{ad.bermudez168@uniandes.edu.co}}
\maketitle
\begin{abstract}
We use the Liouville eigenfunctions to define a classical version
of the geometric tensor and study its relationship with the classical
adiabatic gauge potential (AGP). We focus on integrable systems and
show that the imaginary part of the geometric tensor is related to
the Hannay curvature. The singularities of the geometric tensor and
the AGP allows us to link the transition from Arnold-Liouville integrability
to chaos with some of the mathematical formalism of quantum phase transitions.
\end{abstract}

\section{Introduction}

Classical and quantum mechanics are usually presented in different
mathematical formalisms. However, this does not
need to be the case. Quantum mechanics can be expressed in a Hamiltonian
way \cite{hestlot} and in the same geometrical fashion as classical
mechanics \cite{geometricQM,geometricQM2,geometricQM3}. Equivalently,
classical mechanics can be reformulated as a theory of operators acting
on a Hilbert space \cite{KvN1,KvN2,KvN3,KvN4,KvN5}. This is not to
say that the two theories are equivalent, of course, but that sometimes
mathematical tools developed for one of them can be applied to the
other.

Here we are going to use Liouville eigenfunctions and the Hilbert
space formulation of classical mechanics to construct classical analogs
of two mathematical objects defined in the context of quantum mechanics:
the geometric tensor \cite{quantum tensor,quantum tensor 2} and the
adiabatic gauge potential (AGP)\cite{AGP}. These related objects are of interest for geometric phases \cite{Intro1},  quantum ergodicity \cite{Intro2}, shortcuts to adiabaticity \cite{Intro3}, information geometry \cite{Intro4},  phase transitions \cite{Intro5}, quantum chaos \cite{anatoli}, etc.   With such an extensive field of applications, studying these objects in a classical context seems justified.

We will take a purely classical approach, and no classical limit of a quantum object is ever taken. We focus our attention on integrable systems in the sense of  Arnold-Liouville. For them, we will show that the imaginary part of the geometric tensor is related to the Hannay curvature. We will compare the real part of the tensor with previous results \cite{analog}.

The structure of this paper is as follows: in section 2, we present
the basics of the Hilbert space formulation of classical mechanics,
the so-called Koopman-von Neumann (KvN) theory (see \cite{KvN5} for
a more complete exposition).

In section 3 we derive the classical AGP and show that it is related
to first-order canonical perturbation theory.

In section 4 we define the geometric tensor for classical
states and give an explicit expression for its real and imaginary part
in terms of the AGP.

In section 5 we discuss the relationship between the famous ``small
denominator problem'' of classical perturbation theory, the classical
AGP and the transition into chaos. In this regard, the AGP serves
as a probe for the transition from integrability to chaos in both classical
and quantum mechanics \cite{anatoli}. 

In section 6 we compare our results with the ones obtained in Ref.
\cite{analog}. Although both tensors are shown to be related to the
generating function of first-order perturbation theory, they are different
objects. We use a simple example to compare both formulas.

\section{Hilbert structure over the classical phase space}

Let $\Gamma$ be a $2n$ dimensional phase space. We will use the
shorthand vector notation $\mathbf{q}=\left(q_{1},q_{2},\ldots,q_{n}\right)$
and $\mathbf{p}=\left(p_{1},p_{2},\ldots,p_{n}\right)$. The elements
of the Hilbert space $\mathcal{H}(\Gamma)$ are the square-integrable
phase-space wave functions $\psi(\mathbf{q},\mathbf{p})$,

\begin{equation}
\int_{\varGamma}\left|\psi(\mathbf{q},\mathbf{p})\right|{}^{2}\,\mathrm{d}\mathbf{q}\mathrm{d}\mathbf{p}<\infty,
\end{equation}
where the integration is over all phase space. The inner product in
$\mathcal{H}(\Gamma)$ is given by

\begin{equation}
(\varphi,\psi)=\int_{\varGamma}\varphi^{*}(\mathbf{q},\mathbf{p})\psi(\mathbf{q},\mathbf{p})\,\mathrm{d}\mathbf{q}\mathrm{d}\mathbf{p}.\label{inner}
\end{equation}

There are infinite ways to associate phase space functions with Hermitian
operators acting on $\mathcal{H}(\Gamma)$. In this paper, we are going
to use two of them, and we will refer to them as the multiplicative
and the KvN rules \cite{KvN rule}.

The multiplicative rule is the simplest one, for any phase-space
function $f(\mathbf{q},\mathbf{p})$ we define the operator $\hat{f}$
by
\begin{equation}
\hat{f}\psi(\mathbf{q},\mathbf{p})=f(\mathbf{q},\mathbf{p})\psi(\mathbf{q},\mathbf{p}).
\end{equation}
The most important operators defined this way are the position and
momentum operators

\begin{align}
\hat{q}_{i}\psi(\mathbf{q},\mathbf{p}) & =q_{i}\psi(\mathbf{q},\mathbf{p}),\nonumber \\
\hat{p}_{j}\psi(\mathbf{q},\mathbf{p}) & =p_{j}\psi(\mathbf{q},\mathbf{p}).
\end{align}

On the other hand, the KvN rule is given by
\begin{equation}
\hat{f}\psi(\mathbf{q},\mathbf{p})=-i\{\psi(\mathbf{q},\mathbf{p}),f(\mathbf{q},\mathbf{p})\}.\label{multiplication}
\end{equation}
We can use the KvN rule to define a couple of operators canonically
conjugate to $(\hat{\mathbf{q}},\hat{\mathbf{p}})$, they are

\begin{align*}
\hat{\eta}_{i}\psi(\mathbf{q},\mathbf{p}) & =-i\{\psi(\mathbf{q},\mathbf{p}),p_{i}\}=-i\hbar\frac{\partial}{\partial q_{i}}\psi(\mathbf{q},\mathbf{p}),\\
\hat{\theta}_{j}\psi(\mathbf{q},\mathbf{p}) & =i\{\psi(\mathbf{q},\mathbf{p}),q_{j}\}=-i\hbar\frac{\partial}{\partial p_{j}}\psi(\mathbf{q},\mathbf{p}).
\end{align*}

The operators $(\hat{\mathbf{q}},\hat{\mathbf{p}})$ and $(\hat{\eta},\hat{\theta})$
are conjugate to each other in the quantum sense, this is, they obey
the canonical commutation relations

\begin{align}
\bigl[\hat{q}_{i},\hat{\eta}_{j}\bigr] & =\bigl[\hat{p}_{i},\hat{\theta}_{j}\bigr]=i\hbar\delta_{ij},\nonumber \\
\bigl[\hat{q}_{i},\hat{p}_{j}\bigr] & =\bigl[\hat{q}_{i},\hat{\theta}_{j}\bigr]=\bigl[\hat{p}_{i},\hat{\eta}_{j}\bigr]=\bigl[\hat{\eta}_{i},\hat{\theta}_{j}\bigr]=0.\label{comm}
\end{align}

The time evolution of $\psi(\mathbf{q},\mathbf{p})$ is given by the
Schrodinger-like equation
\[
i\frac{\partial}{\partial t}\psi(\mathbf{q},\mathbf{p})=\hat{L}\psi(\mathbf{q},\mathbf{p}),
\]
where the Liouvillian is 

\begin{align*}
\hat{L} & =-i\{,H(\mathbf{q},\mathbf{p})\}=-i\sum_{j=1}^{n}\left(\frac{\partial H}{\partial p_{j}}\frac{\partial}{\partial q_{j}}-\frac{\partial H}{\partial q_{j}}\frac{\partial}{\partial p_{j}}\right)\\
 & =\sum_{j=1}^{n}\left(\frac{\partial H}{\partial p_{j}}\hat{\eta}_{j}-\frac{\partial H}{\partial q_{j}}\hat{\theta}_{j}\right).
\end{align*}

Finally, notice that the probability density given by the Born rule
$\rho=\bigl|\psi\bigr|^{2}$ obeys the Liouville equation
\[
\frac{\partial}{\partial t}\rho=\{H,\rho\},
\]
and the expectation value of any operator given by the multiplication
rule (\ref{multiplication}) is
\[
(\psi,\hat{f}\psi)=\int_{\varGamma}\rho f\:\mathrm{d}\mathbf{q}\mathrm{d}\mathbf{p}
\]
Hence, the Hilbert space formalism we presented here is equivalent
to classical statistical mechanics.

\subsection{Integrable systems}

For integrable systems in the sense of Arnold-Liouville, we can make
a canonical transformation to angle-action variables $(\mathbf{q},\mathbf{p})\rightarrow(\mathbf{I},\pmb{\phi})$
such that $H(\mathbf{q},\mathbf{p})\rightarrow H(\mathbf{I})$. In
this case, the Liouvillian simplifies to 

\begin{equation}
\hat{L}=-i\{,H(\mathbf{I})\}=-i\pmb{\omega}\cdot\frac{\partial}{\partial\pmb{\phi}}.\label{L}
\end{equation}
The eigenfunctions of (\ref{L}) are \cite{wilkie,wilkie2}

\begin{equation}
\psi_{\mathbf{I}',\mathbf{k}}(\mathbf{I},\pmb{\phi})=\frac{1}{(2\pi)^{n/2}}\delta(\mathbf{I}-\mathbf{I}')e^{i\mathbf{k}\cdot\pmb{\phi}},\label{eigen}
\end{equation}
with $\mathbf{k}\in\mathbb{Z}^{n}$, and the eigenvalues are given
by

\begin{equation}
l_{\mathbf{I}',\mathbf{k}}=\mathbf{k}\cdot\pmb{\omega}(\mathbf{I}').\label{eigenvalue}
\end{equation}
In the above expressions, $\mathbf{I}=\mathbf{I}(\mathbf{q},\mathbf{p})$
is a function of the phase-space coordinates whereas $\mathbf{I}'$
is a vector of positive numbers.

The spectrum of (\ref{L}) is more complex than it looks at first
glance since it depends on the frequencies $\pmb{\omega}(\mathbf{I}')$.
For example, the frequencies of the n-dimensional isotropic Harmonic
oscillator are independent of the action so the spectrum is discrete
and each eigenvalue is continuously degenerate. However, the eigenvalues (\ref{eigenvalue}) form a continuum in most other cases. 

The eigenfunctions (\ref{eigen}) are not normalizable, and their inner product is distribution valued. However, we want our definition
of the geometric tensor to be valued in real numbers. To accomplish this, we are going to understand the orthonormality condition in the
sense of eigendifferentials \cite{greiner}. We first take the interval
$\delta\mathbf{I}'$ and consider the ball $\mathcal{B}_{\mathbf{I}'}$
centered in $\mathbf{I}'$ with radius $\bigl\Vert\delta\mathbf{I}'\bigr\Vert$,
the orthonormality condition for the Liouville eigenfunctions we are
going to use is 
\begin{equation}
(\psi_{\mathbf{I}',\mathbf{N}},\psi_{\mathbf{I}'',\mathbf{M}})=\left\{ \begin{array}{c}
\delta_{\mathbf{M},\mathbf{N}}\qquad\mathrm{if}\:\mathbf{I}''\in\mathcal{B}_{\mathbf{I}'}\\
\:\quad\;0\qquad\mathrm{otherwise}.
\end{array}\right.
\end{equation}
In the above, the distance $\bigl\Vert\delta\mathbf{I}'\bigr\Vert$
can be as small as we want but not zero. When necessary, the inner
products calculated below are taken in the sense of eigendifferential
to obtain real-valued results.

We end this section by pointing out that 
\begin{equation}
\int_{\varGamma}\:\mathrm{d}\mathbf{q}\mathrm{d}\mathbf{p}=\int_{I}\int_{\phi}\:\mathrm{d}\mathbf{I}\mathrm{d}\pmb{\phi},
\end{equation}
where each integral of the actions is evaluated from $0$ to $\infty$,
and the integrals for the angles are evaluated from $0$ to $2\pi$.
We will also use the following notation for the angle average of a
function
\begin{equation}
\left\langle \cdots\right\rangle =\frac{1}{(2\pi)^{n/2}}\int_{\varGamma}\,(\cdots)\,\delta(\mathbf{I}(\mathbf{p},\mathbf{q})-\mathbf{I}')\mathrm{d}\mathbf{q}\mathrm{d}\mathbf{p}=\frac{1}{(2\pi)^{n/2}}\int_{\phi}(\cdots)\,\mathrm{d}\pmb{\phi}.
\end{equation}

\section{Adiabatic gauge potential}

In this section, we generalize in two ways the results obtained in
\cite{adiabaticdriving}. First, we will calculate the AGP for the
case of several degrees of freedom. Additionally, we will work with
the distribution-valued eigenfunctions (\ref{eigen}).

Let us first recall that any well-behaved periodic function in phase
space can be expanded as a Fourier series of the form

\begin{align*}
f(\mathbf{I},\pmb{\phi}) & =\frac{1}{(2\pi)^{n/2}}\sum_{\mathbf{k}}a_{\mathbf{k}}(\mathbf{I})e^{i\mathbf{k}\cdot\pmb{\phi}}.\\
a_{\mathbf{k}} & (\mathbf{I})=\bigl\langle f(\mathbf{I},\mathbf{\phi})e^{-i\mathbf{k}\cdot\pmb{\phi}}\bigr\rangle.
\end{align*}

Let $H(q,p,\lambda)=H(I(\lambda),\lambda)$ be a classical Integrable
family of Hamiltonians with dependency on some external parameters
$\lambda=(\lambda_{1},\lambda_{2},...,\lambda_{n})\in\mathit{\mathcal{M}}$,
and \textit{$\mathcal{M}$ }is a smooth manifold. The parameter dependence
of the Hamiltonian induces a dependence in the Liouvillian $\hat{L}(\lambda)$
and in its eigenfunctions $\psi_{\mathbf{I}',\mathbf{k}}(\lambda)$.
We are going to use $\mathrm{d}_{\lambda}$ to designate the exterior
derivative with respect to parameters $\lambda$.

The AGP $\hat{\mathcal{A}}$ is an Hermitian operator that has the
information about the variation of the eigenfunctions, it is defined
as

\begin{align}
(\psi_{\mathbf{\mathbf{I}'},\mathbf{k}},\hat{\mathcal{A}}\psi_{\mathbf{\mathbf{I}'},\mathbf{k}})= & 0,\label{a}\\
(\psi_{\mathbf{\mathbf{I}'},\mathbf{k'}},\hat{\mathcal{A}}\psi_{\mathbf{\mathbf{I}'},\mathbf{k}})= & -i\frac{(\psi_{\mathbf{\mathbf{I}'},\mathbf{k'}},\mathrm{d}_{\lambda}\hat{L}\psi_{\mathbf{\mathbf{I}'},\mathbf{k}})}{l_{\mathbf{k}}-l_{\mathbf{k'}}}\nonumber \\
= & -i\frac{(\psi_{\mathbf{I}',\mathbf{k'}},\mathrm{d}_{\lambda}\hat{L}\psi_{\mathbf{I}',\mathbf{k}})}{\pmb{\omega}(\mathbf{I}')\cdot(\mathbf{k}-\mathbf{k}')},\label{b}
\end{align}
where (\ref{a}) is the Berry-Simon condition of parallel transport,
and (\ref{b}) gives the variation of the eigenfunctions with respect
to the parameters such that

\begin{equation}
-i\mathrm{d}_{\lambda}\psi_{\mathbf{\mathbf{I}'},\mathbf{k}}=\hat{\mathcal{A}}\psi_{\mathbf{I}',\mathbf{k}}=-i\sum_{m\neq n}\frac{(\psi_{\mathbf{I}',\mathbf{k'}},\mathrm{d}_{\lambda}\hat{L}\psi_{\mathbf{I}',\mathbf{k}})}{\pmb{\omega}\cdot(\mathbf{k}-\mathbf{k}')}\psi_{\mathbf{I}',\mathbf{k'}}.\label{defA}
\end{equation}
Replacing (\ref{L}) and (\ref{eigen}) into (\ref{defA}) and changing
the variable of summation, we get

\begin{equation}
\hat{\mathcal{A}}\psi_{\mathbf{I}',\mathbf{k}}=-ie^{i\mathbf{k}\cdot\pmb{\phi}}\mathbf{\mathbf{k}}\cdot\sum_{\mathbf{k'}\neq\mathbf{0}}\frac{1}{(2\pi)^{n}\pmb{\omega}\cdot\mathbf{k'}}\frac{\partial}{\partial\mathbf{I}}\left\langle \mathrm{d}_{\lambda}He^{-i\mathbf{k'}\cdot\pmb{\phi}}\right\rangle ,\label{Apsi}
\end{equation}
where $\pmb{\omega}$ and $\mathrm{d}_{\lambda}H$ are evaluated at $\mathbf{\mathbf{I}'}$.
Since (\ref{Apsi}) above does not contain the term $\mathbf{k}=\mathbf{0}$,
the Fourier series that appears corresponds to a function that does
not have a secular term. We can then write the following expression 

\begin{equation}
\hat{\mathcal{A}}\psi_{\mathbf{\mathbf{I}'},\mathbf{k}}=-i\{\psi_{\mathbf{I}',\mathbf{k}},W\},\label{AW}
\end{equation}
where the generating function $W$ is given by the Fourier expansion

\begin{align}
W & =i\sum_{\mathbf{k'}}\frac{\mathcal{W}_{\mathbf{k'}}}{\pmb{\omega}\cdot\mathbf{k}'}e^{i\mathbf{k'}\cdot\pmb{\phi}},\nonumber \\
\mathcal{W}_{\mathbf{k}'} & =\bigl\langle(\mathrm{d}_{\lambda}H-\bigl\langle\mathrm{d}_{\lambda}H\bigr\rangle)e^{-i\mathbf{k'}\cdot\pmb{\phi}}\bigr\rangle,\label{saletan}
\end{align}
and the role of $\bigl\langle\mathrm{d}_{\lambda}H\bigr\rangle$ is
to eliminate any secular term so $W$ is completely periodic, i.e.,
$\bigl\langle W\bigr\rangle=0.$ We can then make the identification
\[
\hat{\mathcal{A}}=-i\left\{ ,W\right\} .
\]

The expression (\ref{saletan}) corresponds to the generating function
of first order canonical perturbation theory \cite{Saletan,ferraz}
just as the quantum AGP is related to quantum perturbation theory
\cite{adriving2}. We can see that $\hat{\mathcal{A}}$ is obtained
by the KvN rule applied to $W$.

We give in the appendix an alternative derivation of the AGP without
using the Liouville eigenfunctions nor the Hilbert space formalism
used so far.

\section{The classical geometric tensor}

By analogy with the quantum case, we define the classical geometric
tensor for integrable systems by 

\begin{equation}
Q_{\mu\nu}^{\mathbf{k}}(\mathbf{I}')=(\partial_{\mu}\psi_{\mathbf{I}',\mathbf{k}},\partial_{\nu}\psi_{\mathbf{I}',\mathbf{k}})-(\partial_{\mu}\psi_{\mathbf{I}',\mathbf{k}},\psi_{\mathbf{I}',\mathbf{k}})(\psi_{\mathbf{I}',\mathbf{k}},\partial_{\nu}\psi_{\mathbf{I}',\mathbf{k}}),\label{Gtensor}
\end{equation}
where $\partial_{\mu}=\frac{\partial}{\partial\lambda_{\mu}}$.

The imaginary part of the geometric tensor is 

\begin{equation}
\mathrm{Im\,}Q{}_{\mu\nu}^{\mathbf{k}}(\mathbf{I}')=-\frac{1}{2}F_{\mu\nu}^{\mathbf{k}}(\mathbf{\mathbf{I}}'),\label{imQ0}
\end{equation}
where $F_{\mu\nu}^{\mathbf{m},\mathbf{k}}$ is the Berry curvature

\begin{equation}
F_{\mu\nu}^{\mathbf{k}}(\mathbf{\mathbf{I}}')=i\partial_{\mu}(\partial\psi_{\mathbf{\mathbf{I}'},\mathbf{k}},\partial_{\nu}\psi_{\mathbf{\mathbf{I}'},\mathbf{k}})-i\partial_{\nu}(\partial\psi_{\mathbf{\mathbf{I}'},\mathbf{k}},\partial_{\mu}\psi_{\mathbf{\mathbf{I}'},\mathbf{k}}).
\end{equation}
For integrable systems, we expect that the geometric phase of the
Liouville eigenfunctions to be related to Hannay angles by \cite{adiabaticdriving}

\begin{equation}
\mathrm{Classical\:geometric\:phase}=\int F_{\mu\nu}^{\mathbf{\mathbf{I}'},\mathbf{k}}\delta\lambda_{\mu}\delta\lambda_{\nu}=-\mathbf{k}\cdot\Delta\pmb{\phi}(\mathbf{I}'),\label{imQ1}
\end{equation}
where $\Delta\pmb{\phi}(\mathbf{I}')$ is the vector with components given
by the holonomy for each angle variable. 

On the other hand, 
\[
\mathrm{Re}\,Q_{\mu\nu}^{\mathbf{k}}(\mathbf{I}')=\frac{1}{4}\mathcal{F}_{\mu\nu},
\]
where $\mathcal{F}_{\mu\nu}$ are the components of the ``quantum''
Fisher information matrix \cite{fishermatrix}, a fundamental quantity
in quantum multiparameter estimation and quantum metrology.
\[
\]

We will now use (\ref{AW}) to compute the classical geometric tensor
for integrable systems. For variations obeying (\ref{defA}), we can
rewrite $Q_{\mu\nu}^{\mathbf{k}}(\mathbf{I}')$ in terms of the AGP
as follows
\begin{align*}
Q_{\mu\nu}^{\mathbf{k}}(\mathbf{I}') & =(\partial_{\mu}\psi,\partial_{\nu}\psi)-(\partial_{\mu}\psi,\psi)(\psi,\partial_{\nu}\psi)\\
 & =-(\psi,\hat{\mathcal{A}}_{\mu}\hat{\mathcal{A}}_{\nu}\psi)+\bigl\langle\hat{\mathcal{A}}_{\mu}\psi,\psi)(\psi,\hat{\mathcal{A}}_{\nu}\psi),
\end{align*}
where we omitted the sub-indices for notation simplicity but we are
still working with the eigenfunctions (\ref{eigen}). Using (\ref{AW}),
we can write the tensor as

\begin{equation}
Q_{\mu\nu}^{\mathbf{k}}(\mathbf{I}')=(\psi,\{\{\psi,W_{\nu}\},W_{\mu}\})-(\{\psi,W_{\mu}\},\psi)(\psi,\{\psi,W_{\nu}\}).\label{Ctensor}
\end{equation}
Since $W$ is completely periodic, the terms like $\bigl\langle\psi,\{\psi,W_{\nu}\}\bigr\rangle$
vanishes. Thus, Eq. (\ref{Ctensor}) simplifies to

\begin{equation}
Q_{\mu\nu}^{\mathbf{k}}(\mathbf{I}')=(\psi,\{\{\psi,W_{\nu}\},W_{\mu}\})=(\{\psi,W_{\mu}\},\{\psi,W_{\nu}\}).\label{Ctensor2}
\end{equation}

The real and imaginary parts of the tensor are 

\begin{align}
\mathrm{Re}\,Q_{\mu\nu}^{\mathbf{k}}(\mathbf{I}') & =\bigl\langle\mathbf{k\cdot}\frac{\partial W_{\mu}}{\partial\mathbf{I}}\mathbf{k\cdot}\frac{\partial W_{\nu}}{\partial\mathbf{I}}\bigr\rangle,\label{reQ0}\\
\mathrm{Im\,Q_{\mu\nu}^{\mathbf{k}}(\mathbf{I}')} & =\frac{i}{2}(\psi,\{\{\psi,W_{\nu}\},W_{\mu}\})-\frac{i}{2}(\psi,\{\{\psi,W_{\mu}\},W_{\nu}\})\nonumber \\
 & =\frac{i}{2}(\psi,\{\{\psi,\{W_{\mu},W_{\nu}\}\})=\frac{1}{2}\mathbf{k\cdot}\frac{\partial}{\partial\mathbf{I}}\bigl\langle\{W_{\mu},W_{\nu}\}\bigr\rangle,\label{ImQ}
\end{align}
where in the last line we have used the Jacobi identity for the Poisson
brackets. The Hannay curvature 2-form of the jth angle variable can
be expressed in terms of the generating functions $W$ by \cite{adiabaticdriving}

\[
f_{\mu\nu(j)}=\frac{\partial}{\partial I_{j}}\bigl\langle\{W_{\mu},W_{\nu}\}\bigr\rangle.
\]
Hence, the imaginary part of the classical geometric tensor can be
written as

\begin{equation}
\mathrm{Im\,Q_{\mu\nu}^{\mathbf{k}}(\mathbf{I}')}=\frac{1}{2}\mathbf{k\cdot}\mathbf{f}_{\mu\nu}(\mathbf{I}').\label{IMQ}
\end{equation}
The term $F_{\mu\nu}^{\mathbf{k}}(\mathbf{\mathbf{I}}')=-\mathbf{k\cdot}\mathbf{f}_{\mu\nu}(\mathbf{I}')$
is the classical version of the Berry curvature for the Liouville
eigenfunctions, in agreement with (\ref{imQ0}) and (\ref{imQ1}),
and as expected from the results of \cite{adiabaticdriving}. 

\section{Phase transitions into chaos}

In quantum mechanics, we expect that the geometric tensor shows some singular behavior when a quantum phase transition happens \cite{zanardi 1,zanardi,zanardi2,zanardi4,zanardi5}. The logic behind this intuition is that the geometric tensor defines a distance between states, and we expect that two states $\psi(\lambda)$ and $\psi(\lambda+\delta\lambda)$ should be significantly different if a phase transition occurs within  $(\lambda,\,\lambda+\delta\lambda)$. 
The AGP gives another tool to study the phase transition since its
Hilbert-Smith metric is related to the geometric tensor by \cite{del campo}

\[
\bigl\Vert\delta\lambda\cdot\hat{\mathcal{A}}(\lambda)\bigr\Vert_{\mathrm{HS}}^{2}=\sum_{\mathbf{m},\mathbf{k}}\sum_{\mu,\nu}Q_{\mu\nu}\delta\lambda_{\mu}\delta\lambda_{\nu}.
\]
Therefore, if there are singularities in the geometric tensor,
we can expect a similar behavior in $\hat{\mathcal{A}}$ \cite{hatomura}. 

In the classical integrable cases we have studied, we can see that the singularities in  $Q_{\mu\nu}^{\mathbf{k}}$ and $\hat{\mathcal{A}}$ indeed mark a drastic change in the classical eigenstates. 
Starting from integrable systems, the
classical AGP will have singularities when $\pmb{\omega}\cdot\mathbf{k}'$
vanishes or it is small enough. This is the famous small denominator
problem of canonical perturbation theory.

Let us remember that $\pmb{\omega}$ depends on the value of the actions.
Hence, some eigenfunctions $\psi_{\mathbf{\mathbf{I}'},\mathbf{k}}$ at $\lambda$, the ones belonging to irrational enough tori, can be very close to their equivalent at $\lambda+\delta\lambda$, giving
rise to areas of integrability. However, the possibility of singularities
in the AGP imply that some eigenfunctions of the Liouvillian cannot
be reached via a continuous path of unitary transformations that start
in an integrable state. If the AGP is singular around $(\lambda+\delta\lambda)$,
then there exist some eigenfunctions $\psi(\lambda)$ such that the
following parallel transport equation is not true for any path $\gamma$
in parameter space
\begin{equation}
\psi(\lambda)=U_{\gamma}\psi(\lambda_{0}),
\end{equation}
where $\lambda_{0}$ is outside $(\lambda+\delta\lambda)$, $\psi(\lambda_{0})$
correspond to an integrable state, and $U$ is the path-dependent unitary transformation given by

\begin{equation}
U_{\gamma}=Pe^{i\int_{\gamma}\hat{\mathcal{A}}}.
\end{equation}

However, there is a clear difference between the classical transition
into chaos and quantum phase transition, despite both being related
to divergences of the geometric tensor and the AGP. The primary interest in quantum phase transition is the behavior of the ground state of
the quantum Hamiltonian, meanwhile, the Liouvillian (\ref{L}) does
not even have a ground state.

\section{Comparison with a previous work}

In this section, we compare our approach with the one given by Gonzalez,
Guti\'errez-Ruiz and Vergara \cite{analog}. Suppose we have a family
of integrable Hamiltonians $H(q,p,\lambda)$, and that we can solve
for parameter-dependent action-angle variables $\mathbf{I}(\lambda)$
and $\phi(\lambda)$ such that $H=H(\mathbf{I}(\lambda),\lambda)$.
Since we know the functional form of the action-angle variables, we
can compute their differential and express it as the flow given by
a generating function \cite{tesis}
\begin{align}
I_{j}(\lambda+\delta\lambda)-I_{j}(\lambda) & =\mathrm{d}_{\lambda}I_{j}(\lambda)=\{I_{j}(\lambda),G(\mathbf{I},\pmb{\phi},\lambda)\},\nonumber \\
\phi_{j}(\lambda+\delta\lambda)-\phi_{j}(\lambda) & =\mathrm{d}_{\lambda}\phi_{j}(\lambda)=\{\phi_{j}(\lambda),G(\mathbf{I},\pmb{\phi},\lambda)\}.
\end{align}

As with all functions, we can decompose $G$ into its secular and
its completely periodic parts
\begin{align}
G(\mathbf{I},\pmb{\phi},\lambda) & =W(\mathbf{I},,\pmb{\phi},\lambda)+\alpha(\mathbf{I},\lambda),\nonumber \\
\bigl\langle G\bigr\rangle & =\alpha.\label{G}
\end{align}
By construction, $W$ is the generating function given by first-order
canonical perturbation theory \cite{Saletan}. The formula (\ref{saletan})
is designed to give the variation of the action 
\[
\mathrm{d}_{\lambda}I_{j}(\lambda)=\left\{ I_{j}(\lambda),W(\mathbf{I},,\pmb{\phi},\lambda)\right\} .
\]
We can see that $\alpha(\mathbf{I},\lambda)$ has no effect on $\mathrm{d}_{\lambda}I_{j}(\lambda)$,
it only affects the angle variable. 

Gonzales \emph{et al }define their classical metric tensor (comparable
to the real part of the geometric tensor) by the formula

\begin{align}
g_{\mu\nu} & =\bigl\langle G_{\mu}G_{\nu}\bigr\rangle-\bigl\langle G_{\mu}\bigr\rangle\bigl\langle G_{\nu}\bigr\rangle,\label{G0}
\end{align}
that, in view of (\ref{G}), can also be written as

\begin{equation}
g_{\mu\nu}=\bigl\langle W_{\mu}W_{\nu}\bigr\rangle.\label{Guv}
\end{equation}

We can see that both (\ref{reQ0}) and (\ref{Guv}) are related to
the generating function from first-order canonical perturbation theory,
but beyond that, their relationship is not clear. In the next subsection,
we are going to compare them for a simple system.

\subsection{Example: Generalized oscillator}

To compare results, we are going to calculate the real part of the
geometric tensor (\ref{reQ}) on the system given by the Hamiltonian 

\begin{equation}
H=\frac{1}{2}\left(Xq^{2}+2Yqp+Zp^{2}\right),
\end{equation}
where $X,Y,$ and $Z$ are the parameters to be varied. The canonical
transformation to angle--action variables is \cite{cruchinski}

\begin{align}
q & =\sqrt{\frac{2IZ}{\omega}}\cos\phi,\nonumber \\
p & =-\sqrt{\frac{2IZ}{\omega}}(\frac{Y}{Z}\cos\phi+\frac{\omega}{Z}\sin\phi),\nonumber \\
\omega & =\sqrt{(XZ-Y^{2})},\;\,\:XZ>Y^{2},
\end{align}
leads to the following generating functions \cite{adiabaticdriving,tesis} 

\begin{align}
W_{Y} & =\frac{ZI}{2\omega^{2}}(\frac{Y}{Z}\sin2\phi-\frac{\omega}{Z}\cos2\phi),\nonumber \\
W_{X} & =-\frac{ZI}{4\omega^{2}}\sin(2\phi),\nonumber \\
W_{Z} & =-\frac{ZI}{4\omega^{2}}(\left[\tfrac{Y}{Z}\right]^{2}\sin2\phi-\left[\tfrac{\omega}{Z}\right]^{2}\sin2\phi-2\tfrac{Y\omega}{Z^{2}}\cos2\phi).\label{Ws}
\end{align}

The Hannay curvature of this system was calculated from (\ref{Ws})
in \cite{adiabaticdriving}. Applying formula (\ref{reQ0}) to the
generators (\ref{Ws}) gives

\begin{equation}
\mathrm{Re}\,Q_{\mu\nu}^{k}=\frac{k^{2}}{32\omega^{4}}\left(\begin{array}{ccc}
Z^{2} & -2XY & 2Y^{2}-XZ\\
-2XY & 4XZ & -2XY\\
2Y^{2}-XZ & -2XY & X^{2}
\end{array}\right).\label{reQ}
\end{equation}

Comparing (\ref{reQ}) with the metric tensor calculated \cite{analog}
for the same system, we see that they are related by

\[
\mathrm{Re}\,Q_{\mu\nu}^{k}=\frac{k^{2}}{I}g_{\mu\nu},
\]
this is, they are proportional to each other in a parameter-independent
way. These two quantities in general may always behave similarly, but the author does not see a way to prove
it at the moment. 

\section{Final remarks. }

The Hilbert space formulation of classical mechanics is full of non-measurable
quantities. Classically, only phase space functions are physical quantities.
Hence, operators like $(\hat{\eta},\hat{\theta})$ or even the Liouvillian
$\hat{L}$ are typically considered hidden variables (see the discussion
about superselection rules in classical mechanics given in \cite{KvN5}).

Our definition of the geometric tensor (\ref{Gtensor}) has the uncomfortable
feature of depending on the vector of integers $\mathbf{k}$, which
is not a function of phase space quantities. Taking into account that
only the probability density $\rho$ and the expectation values calculated
from it have physical meaning, the tensor (\ref{Gtensor}) gives different
results for physically equivalent states (all the states $\psi_{\mathbf{I}',\mathbf{k}}$
with the same $\mathbf{I}'$ differ only by a phase). However, it
is $\mathbf{k}$ that relates the AGP with the transition to chaos
so it is not a clear-cut situation.

On the positive side, our tensor immediately contains the information
about the Hannay curvature, just as the imaginary part of the quantum
tensor is related to the Berry curvature. 

It is somewhat harder to give an interpretation of the real part of
the classical geometric tensor, the Fisher information matrix. At the moment, the author can only
speculate with possible applications to classical statistical mechanics
(see section 2.5 of \cite{fishermatrix} and references therein).

Another bonus of our approach from Liouville eigenfunctions is that
our definition (\ref{Gtensor}) is not restricted to integrable systems.
Liouville eigenfunctions for chaotic systems are complicated \cite{wilkie,wilkie2}
but, in principle, nothing prohibits us from taking the same
approach with them. After all, the inner product between KvN states
(\ref{inner}) is independent of the dynamics.

The similarity between the emergence of Hamiltonian chaos and phase transitions is suggestive, but the results are too preliminary to know whether this is a coincidence or there is some deep connection between these two seemingly unrelated concepts. We will leave the investigation on this issue, the real part of the geometric tensor, and its extension for chaotic systems for future work.

\subsection*{Appendix}

Let us define the operator-valued 1-form

\begin{equation}
\hat{\mathcal{A}}=-i\left\{ ,W\right\} .\label{Aapendix}
\end{equation}
The Lie transform generated by (\ref{Aapendix}) is 

\begin{equation}
U=e^{i\hat{\mathcal{A}}}\approx1+\left\{ ,W\right\} .
\end{equation}

By the theory of canonical transformations, the following identity
is obeyed by all phase-space functions 

\begin{equation}
Uf(\mathbf{I},\pmb{\phi})=f(U\mathbf{I},U\pmb{\phi}).
\end{equation}

We now demand that $U$ have the following property
\begin{equation}
Uf(\mathbf{I}(\lambda),\pmb{\phi}(\lambda))=f(\mathbf{I}(\lambda+\delta\lambda),\pmb{\phi}(\lambda+\delta\lambda)).
\end{equation}
The above implies that 
\begin{align}
U\mathbf{I}(\lambda) & =\mathbf{I}(\lambda+\delta\lambda)\approx\mathbf{I}(\lambda)+\left\{ \mathbf{I},W\right\} ,\nonumber \\
U\pmb{\phi}(\lambda) & =\pmb{\phi}(\lambda+\delta\lambda)\approx\pmb{\phi}(\lambda)+\left\{\pmb{\phi},W\right\} .
\end{align}

As mentioned in section 6, $W$ given by (\ref{saletan}) gives the
correct first-order variation of the action variable according to
canonical perturbation theory. A secular term can be added to the
generating function, but this would only affect the angle variable.
We can dismiss the secular term on the basis that only a completely
periodic function leads to the Berry-Simon condition for parallel
transport (\ref{a}). 

Notice that a completely periodic choice for the generating functions
leads to 
\begin{equation}
\left\langle\pmb{\phi}(\lambda+\delta\lambda)-\pmb{\phi}(\lambda)\right\rangle =0,
\end{equation}
which can also be taken as a parallel transport condition, and it
has the advantage that it is formulated without the use of the Liouville
eigenfunctions.

On the downside, it is not obvious how to generalize this last approach to the chaotic case, whereas a Hilbert space approach seems possible for chaotic eigenfunctions.

\end{document}